\pdfoutput=1
\documentclass[
conference,
  letterpaper,
]{IEEEtran}
\IEEEoverridecommandlockouts

\ifCLASSOPTIONcompsoc
  \usepackage[nocompress]{cite}
\else
    \usepackage{cite}
  \fi

\usepackage[utf8x]{inputenc}
\usepackage{amsmath,amssymb,amsfonts}
\usepackage[operators]{cryptocode}
\usepackage{array}

\ifCLASSOPTIONcompsoc
  \usepackage[caption=false,font=footnotesize,labelfont=sf,textfont=sf]{subfig}
\else
  \usepackage[caption=false,font=footnotesize]{subfig}
\fi

\usepackage{url}
\usepackage{balance}                
\usepackage[capitalise]{cleveref}  
\usepackage{booktabs}              
\usepackage{bm}                    
\usepackage{siunitx}               
\usepackage{tikz, adjustbox, tikz-qtree}

\usepackage{algorithm}
\usepackage{algpseudocode}         
\usepackage[protrusion=false]{microtype}

\usepackage{xcolor}

\newcommand{\squeezeup}{\vspace{-.85\baselineskip}}
\microtypesetup{
    protrusion=false,
    final,
    tracking=false,
    kerning=true,
    spacing=true,
    factor=1100, shrink=30}
\SetTracking{encoding={*}, shape=sc}{40}
\usepackage[subtle]{savetrees}

\DeclareSIUnit\px{px}

\Crefname{figure}{Fig.}{Fig.}

\usepackage[super]{nth}

\usepackage{adjustbox}

\usepackage[status=final, mode=multiuser, layout=margin]{fixme}
\fxsetface{margin}{\linespread{.6}\tiny}

\FXRegisterAuthor{kn}{akn}{\tiny\colorbox{EricssonRed}{\color{white}Karl}}
\FXRegisterAuthor{mi}{ami}{\tiny\colorbox{EricssonBlue}{\color{white}Martin}}
\newcommand{\knote}[1]{\knnote{#1}}
\newcommand{\mnote}[1]{\minote{#1}}

\newcommand{\flconstfnt}[1]{\ensuremath{\mathsf{#1}}}
\newcommand{\fltermdeffnt}[1]{{\emph{#1}}}
\newcommand{\R}{\mathbb{R}}
\newcommand{\BigO}[1]{\(\mathcal{O}\left(#1\right)\)}

\definecolor{EricssonBlue}{RGB}{0,130,240}
\definecolor{EricssonRed}{RGB}{255,50,50 }
\definecolor{EricssonOrange}{RGB}{255,140,10}
\definecolor{EricssonYellow}{RGB}{250,210,45}
\definecolor{EricssonGreen}{RGB}{15,195,115}
\definecolor{EricssonPurple}{RGB}{175,120,210}

\definecolor{EricssonBlack}{RGB}{24,24,24}
\definecolor{EricssonGray1}{RGB}{36,36,36}
\definecolor{EricssonGray2}{RGB}{118,118,118}
\definecolor{EricssonGray3}{RGB}{160,160,160}
\definecolor{EricssonGray4}{RGB}{224,224,224}
\definecolor{EricssonGray5}{RGB}{242,242,242}
\definecolor{EricssonWhite}{RGB}{255,255,255}

\usepackage[acronym,toc,shortcuts]{glossaries}
\glsdisablehyper  

\hypersetup{
    final,
    colorlinks=true, breaklinks=true, bookmarks=false,bookmarksnumbered,
    pdftitle={Secure Federated Learning in 5G Mobile Networks}, 
    pdfauthor={Martin Isaksson, Karl Norrman}, 
    pdfkeywords={}, 
    pdfcreator={PdfLaTeX}, 
    pdfproducer={Pandoc, XeLaTeX with hyperref} 
}
\makeatletter
\g@addto@macro\UrlBreaks{\do\-\do/}
\makeatother

\newcommand{\manuscript}{paper}

\DeclareMathOperator{\Pos}{Pos} 
\DeclareMathOperator{\Sig}{SIG} 
\DeclareMathOperator{\diffie}{DH} 
\DeclareMathOperator{\Mac}{MAC} 

\newcommand{\listBase}[2]{\ensuremath{\mathcal{L}_{#1}^{#2}}}
\newcommand{\selList}{\listBase{\text{S}}{}} 

\newcommand{\maskVec}[2]{\bm{m}_{(#1, #2)}} 
\newcommand{\mKI}{\maskVec{k}{i}}
\newcommand{\mIK}{\maskVec{i}{k}}

\newcommand{\mUdt}[2]{\bm{w}_{(#1, #2)}} 
\newcommand{\mLocUdtkt}{\mUdt{k}{t}} 

\newcommand{\tNF}[1]{\text{NF}_{#1}} 
\newcommand{\pNF}[1]{\Pos{\left(\tNF{#1}\right)}} 

\newacronym{3GPP}{3GPP}{Third Generation Partnership Project}
\newacronym{ANN}{ANN}{Artificial Neural Network}
\newacronym{API}{API}{Application Programming Interface}
\newacronym{CA}{CA}{Certificate Authority}
\newacronym{CNN}{CNN}{Convolutional Neural Network}
\newacronym{DEF}{DEF}{Data Expansion Factor}
\newacronym{DoS}{DoS}{Denial-of-service}
\newacronym{ECIES}{ECIES}{Elliptic Curve Integrated Encryption Scheme}
\newacronym{FEMNIST}{FEMNIST}{Federated Extended MNIST}
\newacronym{FL}{FL}{Federated Learning}
\newacronym{FQDN}{FQDN}{Fully Qualified Domain Name}
\newacronym{GPU}{GPU}{Graphics Processing Unit}
\newacronym{IID}{IID}{Independent and identically distributed}
\newacronym{IP}{IP}{Internet Protocol}
\newacronym{KDF}{KDF}{Key Derivation Function}
\newacronym{KPI}{KPI}{Key Performance Indicator}
\newacronym{KTH}{KTH}{KTH Royal Institute of Technology}
\newacronym{MAC}{MAC}{Message Authentication Code}
\newacronym{ML}{ML}{Machine Learning}
\newacronym{MPC}{MPC}{Multi-Party Computation}
\newacronym{PKI}{PKI}{Public Key Infrastructure}
\newacronym{PLMN}{PLMN}{Public Land Mobile Network}
\newacronym{PRF}{PRF}{Pseudo Random Function}
\newacronym{PRG}{PRG}{Pseudo Random Generator}
\newacronym{PS}{PS}{Public Safety}
\newacronym{PRNG}{PRNG}{Pseudo Random Number Generator}
\newacronym{QoE}{QoE}{Quality of Experience}
\newacronym{RAN}{RAN}{Radio Access Network}
\newacronym{RSA}{RSA}{Rivest-Shamir-Adleman}
\newacronym{RSRP}{RSRP}{Reference Signal Received Power}
\newacronym{SAR}{SAR}{Search and Rescue}
\newacronym{TCP}{TCP}{Transmission Control Protocol}
\newacronym{TLS}{TLS}{Transport Layer Security}
\newacronym{UDP}{UDP}{User Datagram Protocol}
\newacronym{UE}{UE}{User Equipment}
\newacronym{UPF}{UPF}{User Plane Function}
\newacronym{UAV}{UAV}{Unmanned Aerial Vehicle}
\newacronym{USV}{USV}{Unmanned Surface Vehicle}
\newacronym{UUID}{UUID}{Universally Unique Identifier}
\newacronym{WARA-PS}{WARA-PS}{WASP Autonomous Research Arenas --- Public Safety}
\newacronym{WARA}{WARA}{WASP Autonomous Research Arenas}
\newacronym{WASP}{WASP}{Wallenberg AI, Autonomous Systems and Software Program}

\newacronym{AF}{AF}{Application Function}
\newacronym{NF}{\text{NF}}{Network Function}
\newacronym{NRF}{NRF}{NF Repository Function}
\newacronym{NWDA}{NWDA}{Network Data Analytics}
\newacronym{NWDAF}{NWDAF}{Network Data Analytics Function}
\newacronym{AMF}{AMF}{Access and Mobility Management Function}
\newacronym{AUSF}{AUSF}{AUthentication Server Function}
\newacronym{OAM}{OAM}{Operation, Administration, and Maintenance}
\newacronym{SBA}{SBA}{Service Based Architecture}

\newglossaryentry{serviceProducer} {
    first = {\emph{service producer}},
    name = {service producer},
    plural = {service producer},
    description = {A Network Function that offers a service to other Network
                   Functions}
}
\newglossaryentry{serviceConsumer} {
    first = {\emph{service consumer}},
    name = {service consumer},
    plural = {service consumers},
    description = {A Network Function that requests service from another
                   Network Function}
}
\newglossaryentry{nodeHierarchy} {
    first = {\fltermdeffnt{node hierarchy}},
    name = {node hierarchy},
    plural = {node hierarchies},
    symbol = \flconstfnt{NH},
    description = {A set of nodes ordered in a tree structure}
}
\newglossaryentry{modelHierarchy} {
    first = {\fltermdeffnt{model hierarchy}},
    name = {model hierarchy},
    plural = {model hierarchies},
    symbol = \flconstfnt{MH},
    description = {A set of models ordered in a tree structure}
}

\usepackage{enumitem,amssymb}
\newlist{todolist}{itemize}{2}
\setlist[todolist]{label=$\square$}
\usepackage{pifont}
\RequirePackage[normalem]{ulem} 
\RequirePackage{color}\definecolor{RED}{rgb}{1,0,0}\definecolor{BLUE}{rgb}{0,0,1} 

\usepackage{pgffor}
\usepackage{pgfplots}

\pgfkeys{/pgf/number format/.cd,1000 sep={\,}}
\pgfplotsset{compat=newest,
    width=6cm,
    height=3cm,
    scale only axis=true,
    max space between ticks=25pt,
    try min ticks=5,
    every tick label/.append style={font=\large},
    every x tick scale label/.style={at={(xticklabel cs:1)},anchor=north west},
    yticklabel style={/pgf/number format/fixed},
    legend style={cells={align=left}},
    every axis/.style={
        axis y line=left,
        axis x line=bottom,
        axis line style={thick,->,>=latex, shorten >=-.4cm}
    },
    every axis plot/.append style={thick},
    tick style={black, thick}
}
\tikzset{
    semithick/.style={line width=1.2pt},
}

\usetikzlibrary{shapes, decorations, calc, arrows}
\usetikzlibrary{3d,fit,backgrounds, decorations.text}
\usetikzlibrary{positioning, shapes.symbols}
\usetikzlibrary{decorations.pathreplacing, calligraphy}

\tikzset{>=latex}

\usepackage{xparse}

\ExplSyntaxOn
\NewDocumentCommand { \xifnum } { }
    {
        \fp_compare:nTF
    }
\ExplSyntaxOff

\def\colorname{{"EricssonBlue","EricssonRed","EricssonOrange","EricssonYellow","EricssonGreen","EricssonPurple","EricssonGray2","EricssonGray4","EricssonBlack"}}

\newcommand{\sharedkey}[2]{%
    \raisebox{-.5 ex}{\tikz{%
    \pgfmathparse{\colorname[#1]};\edef\prevcolor{\pgfmathresult}
    \pgfmathparse{\colorname[#2]};\edef\nextcolor{\pgfmathresult}
    \draw[fill=\prevcolor, draw=white] (0ex,0) arc(90:270:1ex) -- cycle;
    \draw[fill=\nextcolor, draw=white] (0ex,0) arc(90:-90:1ex) -- cycle; }}}

\usepackage{enumitem}

\providecommand{\tightlist}{%
\setlength{\itemsep}{0pt}\setlength{\parskip}{0pt}\setlength{\parindent}{0pt}}

\let\tightlist\relax
\def\tightlist{}

\newcommand\copyrighttext{%
  \footnotesize \textcopyright 2020 IEEE. Personal use of this material is permitted. Permission from IEEE must be obtained for all other uses, in any current or future media, including reprinting/republishing this material for advertising or promotional purposes, creating new collective works, for resale or redistribution to servers or lists, or reuse of any copyrighted component of this work in other works.}

\newcommand\copyrightnotice{%
\begin{tikzpicture}[remember picture,overlay]
\node[anchor=south,yshift=10pt] at (current page.south) {\fbox{\parbox{\dimexpr\textwidth-\fboxsep-\fboxrule\relax}{\copyrighttext}}};
\end{tikzpicture}%
}

\providecommand{\tightlist}{%
  \setlength{\itemsep}{0pt}\setlength{\parskip}{0pt}\setlength{\parindent}{0pt}}

\begin{document}

    \title{Secure Federated Learning in 5G Mobile Networks}

\author{
\IEEEauthorblockN{
  Martin Isaksson\IEEEauthorrefmark{1}\IEEEauthorrefmark{3}%
  ,
  Karl Norrman\IEEEauthorrefmark{2}\IEEEauthorrefmark{3}%
  }

\IEEEauthorblockA{\IEEEauthorrefmark{1}Ericsson Research, Artificial Intelligence, Torshamnsgatan 21, SE-164
83, Stockholm, Sweden}
\IEEEauthorblockA{\IEEEauthorrefmark{2}Ericsson Security Research, Torshamnsgatan 21, SE-164 83, Stockholm,
Sweden}
\IEEEauthorblockA{\IEEEauthorrefmark{3}KTH Royal Institute of Technology, School of Electrical Engineering and
Computer Science\\
Email: \{martisak, knorrman\}@kth.se}
}
\maketitle

    \begin{abstract}
        \acrfull{ML} is an important enabler for optimizing, securing and
        managing mobile networks. This leads to increased collection and
        processing of data from network functions, which in turn may increase
        threats to sensitive end-user information. Consequently, mechanisms to
        reduce threats to end-user privacy are needed to take full advantage of
        \acrshort{ML}. We seamlessly integrate \acrfull{FL} into the 3GPP 5G
        \acrfull{NWDA} architecture, and add a \acrfull{MPC} protocol for
        protecting the confidentiality of local updates. We evaluate the
        protocol and find that it has much lower communication overhead than
        previous work, without affecting \acrshort{ML} performance.
    \end{abstract}


\glsresetall
\bstctlcite{IEEEexample:BSTcontrol}

\hypertarget{introduction}{%
\section{Introduction}\label{introduction}}

\copyrightnotice

The \nth{5} generation of mobile network technologies, 5G, defines a new
standard for \gls{RAN} allowing billions of connected devices to
transmit more data than ever before. Due to the huge amount of data and
devices, complexity of configuring, managing and securing networks
increase. To meet these new demands, \gls{ML} is an important
enabler~\cite{ericssonaiwhitepaper}.

In turn, data collection is necessary for \gls{ML}. While data
collection leads to insights benefiting system optimization, it can be
sensitive in a privacy and a business sense, and may be used for
nefarious purposes. It is important to respect end users' privacy as
well as protecting business sensitive information by considering privacy
during the entire network lifecycle~\cite{ericssonprivacywhitepaper}.

\hypertarget{motivation}{%
\subsection{Motivation}\label{motivation}}

The telecom industry is now considering collaborative \gls{ML} to
improve privacy when using \gls{ML} for network optimization,
time-series forecasting~\cite{DiazGonzalez1334598}, predictive
maintenance and \acrshort{QoE}
modeling~\cite{DBLP:journals/corr/abs-1906-09248,EricssonReviewFL}.

Collaborative \gls{ML} such as \gls{FL}~proposed for mobile networks in
\cite{DBLP:conf/aistats/McMahanMRHA17,DBLP:journals/corr/abs-1902-01046}
avoids central collection of data and instead perform training of an
\gls{ML} model locally where the data is generated. The local updates
generated by, e.g., base stations as in~\cite{DiazGonzalez1334598}, are
then aggregated by a parameter server into a new global model. When
\gls{FL} is used, attackers are therefore required to compromise each
data generating client individually to obtain its raw data. In this
\manuscript{} we consider precisely these types of 5G use-cases based on
a single operator using \gls{FL} in its own network.

Details of how to realize collaborative learning in 5G on a system
architecture and protocol level has not been investigated. Our scheme is
therefore relevant for industry and is timely as input for
pre-standardization research.

\hypertarget{assets-and-threats}{%
\subsection{Assets and Threats}\label{assets-and-threats}}

We consider the problem of protecting raw data collected by \glspl{NF}
in 5G mobile networks for the purpose of tuning their performance. An
\gls{NF} can be deployed at a base station, which is a rich source of
data that can be used to infer network-wide
insights~\cite{ericssonaiwhitepaper}. Confidentiality of this data could
be lost if an \gls{NF} is compromised. If the data is sent to a central
server such as \gls{NWDAF} for \gls{ML} model training, there is
therefore an increased threat that for example mobility patterns of an
end-user leaks. Collecting data also increases the risk that business
sensitive information leaks, such as system parameters. Consequently,
the central server is a more attractive target because it collects data
from multiple \glspl{NF}.

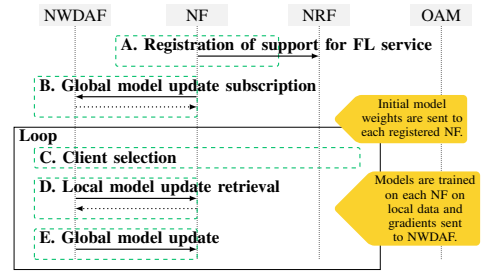
\begin{figure}[t!]
    \centering
    \begin{adjustbox}{width=.70\linewidth}%
        \begin{tikzpicture}[node distance=3cm,
            every node/.style={font=\large},
            comment/.style={
                    rounded corners=1mm, fill=EricssonYellow,
                    font=\sffamily\small\linespread{0.8}\selectfont,
                    inner sep=1mm,%
                    outer sep=1mm,%
                    minimum height=5mm,%
                    align=center,
                    text width=2.5cm,
                    font=\small},
            comment-right/.style={comment, right=.2, signal, signal to=west },
            comment-left/.style={comment, left=.2, signal, signal to=east}]

    \tikzstyle{object} = [rectangle, solid, minimum width=1.5cm, minimum height=.5cm,text centered, fill=EricssonGray5]
    \tikzstyle{lifeline} = [densely dotted, draw=black]
    \tikzstyle{phase} = [dashed, thick, draw=EricssonGreen, rounded corners=.5mm]
    \tikzstyle{arrow} = [thick,->,>=latex]
    \tikzstyle{message} = [outer sep=0pt, inner sep=0pt]
    \tikzstyle{return} = [dotted]
    \tikzstyle{label} = [above, midway]
    \tikzstyle{phaselabel} = [below=.25cm, right, font=\bfseries\large]

    \draw node (NWDAF) [object] {NWDAF} [lifeline] (NWDAF) -- ++(0,-6);
    \draw node (NF) [object, right of=NWDAF] {NF} [lifeline] (NF) -- ++(0,-6);
    \draw node (NRF) [object, right of=NF] {NRF} [lifeline] (NRF) -- ++(0,-6);
    \draw node (OAM) [object, right of=NRF] {OAM} [lifeline]  (OAM) -- ++(0,-6);


    \draw node [phaselabel] at  (1, -.5) {A. Registration of support for FL service};
    \draw [phase] (1, -.5) rectangle (5, -1.25);

    \draw node [message] (in1) at (NF |- 10,-1) {} [arrow] (in1) -- (in1 -| NRF) node [label] {};

    \draw node [phaselabel] at  (-1, -1.5) {B. Global model update subscription};
    \draw [phase] (-1, -1.5) rectangle (3, -2.5);

    \draw node [message] (in2) at (NF |- 10,-2) {} [arrow] (in2) -- (in2 -| NWDAF) node [label] {};
    \draw node [message] (in3) at (NWDAF |- 10,-2.25) {} [arrow, return] (in3) -- (in3 -| NF) node [label] {};

    \draw [draw=black] (-1.5, -2.75) rectangle (7.5, -6.25);
    \draw node [phaselabel] at  (-1.5, -2.75) {Loop};

    \draw node [phaselabel] at  (-1, -3.25) {C. Client selection};
    \draw [phase] (-1, -3.25) rectangle (7, -3.75);

    \draw [phase] (-1, -4) rectangle (3, -5);
    \draw node [phaselabel] at (-1, -4) {D. Local model update retrieval};

    \draw node [message] (in4) at (NWDAF |- 10,-4.5) {} [arrow] (in4) -- (in4 -| NF) node [label] {};
    \draw node [message] (in5) at (NF |- 10,-4.75) {} [arrow, return] (in5) -- (in5 -| NWDAF) node [label] {};

    \node [comment-right] at (6, -4.75) {Models are trained on each NF on local data and gradients sent to NWDAF.};
    \node [comment-right] at (6, -2.55) {Initial model weights are sent to each registered NF.};

    \draw [phase] (-1, -6) rectangle (3, -5.25);
    \draw node [phaselabel] at (-1, -5.25) {E. Global model update};

    \draw node [message] (in7) at (NWDAF |- 10,-5.75) {} [arrow] (in7) -- (in7 -| NF) node [label] {};

\end{tikzpicture}%
    \end{adjustbox}%
    \caption{An overview of our integration of \acrfull{FL} in a 5G \acrfull{NWDA} context.}
    \label{fig:flsequencediagram}
    \squeezeup
\end{figure}

\hypertarget{our-main-contributions}{%
\subsection{Our main contributions}\label{our-main-contributions}}

We design and analyze a scheme which ensures that updates from \gls{FL}
clients are aggregated in a privacy preserving fashion using \gls{MPC}.
We show how this scheme can be integrated in the 5G \gls{NWDA}
framework~\cite{3gpp.23.501,3gpp.23.288} and protocols without breaking
the structure of the architecture or its underlying principles.
Specifically, we consider the following to be our main contributions:

\begin{enumerate}
\def\labelenumi{\arabic{enumi}.}
\tightlist
\item
  An integration of collaborative learning (in particular
  \acrfull{FL}~\cite{DBLP:conf/aistats/McMahanMRHA17}) into the 5G
  \gls{NWDA} framework;
\item
  A privacy-enhancing and efficient scheme for collaborative learning
  algorithms in the 5G \gls{NWDA} inspired by~Bonawitz
  \emph{et al.}~\cite{Bonawitz};
\item
  An evaluation of the scheme with respect to communication cost,
  storage cost and computational cost.
\end{enumerate}

\hypertarget{background}{%
\section{Background}\label{background}}

\hypertarget{gpp-5g-service-based-architecture}{%
\subsection{3GPP 5G Service Based
Architecture}\label{gpp-5g-service-based-architecture}}

The \acrshort{3GPP} created a new framework for core network protocols for 5G
called \gls{SBA}~\cite{3gpp.29.500}.
%
%

\subsubsection{Architecture and Principles}
\gls{SBA} comprises \glspl{NF} that expose
services through RESTful \acrshortpl{API}~\cite{3gpp.29.501}.
\glspl{NF} can invoke services in other \glspl{NF} via these \acrshortpl{API}.
To be discoverable to \glspl{serviceConsumer}, \glspl{serviceProducer} must
register with an \gls{NRF}~\cite{3gpp.29.510}. Upon request from an \gls{NF},
the \gls{NRF} responds with a list of identifiers for suitable
service producers, which can fulfill the service criteria
posed by the \gls{NF}. The \gls{NF} may for example request a list of
all service producers of a certain type.

\knote{One of our main claims is that we seamlessly integrate our solution in
    5G NWDA and 5G SBA. Still, we do not show how we do that, and give
    no justification for the claim. I believe we need to expand on what this 5G
architecture and principles are, and then show our integration.}
%

\subsubsection{Security}
\gls{SBA} builds security in from the start~\cite{3gpp.33.501} so that access
to any \acrshort{API} of an \gls{NF} is authenticated and authorized.
Sensitive data transmitted between providers and consumers is further
confidentiality and integrity protected.

Operators control a \gls{PKI} managing certificates for all \glspl{NF}.\mnote{Reference?}
Our scheme makes use of this \gls{PKI} and does not depend on any of the
authorization features and we assume that all requests are authorized
according to an appropriate policy.

\hypertarget{gpp-5g-analytics-framework}{%
\subsection{3GPP 5G Analytics
framework}\label{gpp-5g-analytics-framework}}

For analytics and predictions, 5G provides harmonized mechanisms
for data collection. 
These mechanisms are based on a consumer and producer concept
realized by \gls{SBA} communication patterns and form the framework called
\gls{NWDA}~\cite{3gpp.23.288}. \gls{NWDA} is centered around a function
named \gls{NWDAF} that serves two main purposes.
First, it acts as a service consumer, collecting data
using \emph{Data Collection} procedures from \glspl{NF}, who act
as service providers.
Second, it processes the data and provides analytics and predictions
as a service provider to other \glspl{NF} using
\emph{Analytics and Prediction Exposure} procedures.

\hypertarget{improving-privacy-with-collaborative-learning}{%
\subsection{Improving privacy with collaborative
learning}\label{improving-privacy-with-collaborative-learning}}

Use-cases relevant for 5G, listed in~\Cref{motivation}, encompasses
\gls{FL} tasks such as time-series forecasting for predictive
maintenance, or classification for traffic management and \acrshort{QoE}
modeling. These tasks revolve around a neural network model that is
parameterized by weights. We therefore consider neural networks, with
weights serialized into a vector \(\bm{w} \in \R^d\), where \(d\) is the
length of the vector.

In geographically distributed collaborative learning frameworks, the
data belongs to and is local to each \gls{NF}.

The collaborative learning approach
\gls{FL}~\cite{DBLP:conf/aistats/McMahanMRHA17} minimizes objective
functions over a set of geographically distributed clients. Training is
done synchronously, and a parameter server coordinates the training
during a number of training rounds. In training round \(t\), each
\(\tNF{k}\) will train locally on~\(n_k\) samples before a local model
update \(\mLocUdtkt\)\mnote{$\bm{w} \neq  \mLocUdtkt$} is sent to the
\gls{NWDAF}. The \gls{NWDAF} performs aggregation of all received local
updates and distributes the global model to all \glspl{NF}.

\hypertarget{integrating-federated-learning-in-the-analytics-framework}{%
\section{Integrating Federated Learning in the Analytics
Framework}\label{integrating-federated-learning-in-the-analytics-framework}}

\label{sect:fl-in-nwda} In this section we present our first main
contribution~--- an integration of collaborative learning, specifically
\gls{FL}, into the 5G \gls{NWDA} framework.

The \glspl{NF} trust the \gls{NWDAF} to faithfully compute the joint
model update of the participating \glspl{NF} individual updates. They
would however like to reduce the risk that the \gls{NWDAF} gets direct
access to the raw local data.

The \gls{NWDAF} may try to insert, delete or modify messages sent
between \glspl{NF}. Because \gls{SBA} provides confidentiality and
integrity protection of messages, \glspl{NF} or other parties cannot
interfere or eavesdrop on the communication between \glspl{NF} and the
\gls{NWDAF}. This is the trust model we use.

To integrate \gls{FL} into the \gls{NWDA} framework, we first split it
into five phases, as seen in \Cref{fig:flsequencediagram}. We consider
the \emph{client} from~\cite{DBLP:conf/aistats/McMahanMRHA17} as a
component of an \gls{NF} and the parameter \emph{server} as a component
of a \gls{NWDAF}.

\hypertarget{registration-of-support-for-service}{%
\subsection{\texorpdfstring{Registration of support for \acrlong{FL}
service}{Registration of support for  service}}\label{registration-of-support-for-service}}

An \gls{NF} may provide the \gls{SBA} service to train an \gls{ML} model
on local data and send model updates to a subscriber of that service.
The model updates are sent as events using the \emph{Data Collection}
procedures defined in~\cite{3gpp.23.288}. We refer to these updates as
\emph{local model updates}. The \gls{NWDAF} is the intended consumer for
these services.

During the registration of the \gls{ML} training \gls{SBA} service, an
\gls{NF} informs the \gls{NRF} about supported features related to model
training. The \gls{NF} can for example indicate the current traffic
load, if it has a \gls{GPU} or which types of models it can train.

\hypertarget{global-model-update-subscription}{%
\subsection{Global model update
subscription}\label{global-model-update-subscription}}

An \gls{NWDAF} provides the \gls{NWDA} analytics service of sending
global model updates to \glspl{NF}. \glspl{NF} interested in global
model updates need to subscribe to them. It is not required that a
consuming \gls{NF} also makes itself available as a training service
provider, but we will assume that this is the case. Therefore,
\glspl{NF} that have subscribed to global model updates, \(K\) such
clients, are available for selection.

\hypertarget{client-selection}{%
\subsection{Client selection}\label{client-selection}}

A necessary part of \gls{FL} is to select a subset of clients in each
training round. It is common to use a random subset of all clients. This
can be suboptimal in the context of 5G where it can, for example,
introduce bias and
unfairness~\cite{DBLP:journals/corr/abs-1912-04977,ericssonaiwhitepaper}.

\gls{FL} algorithms define parameters controlling their behaviors, such
as the fraction \(C\) of \glspl{NF} selected for training in each round.
We denote the set of selected \glspl{NF} \(\selList\) and the number of
selected \glspl{NF}
~\(K_{s} = \vert{}\selList\vert{} = \left\lceil CK \right\rceil\) and
note that the selection strategy does not impact the security of our
scheme. In each training round, the \gls{NWDAF} performs client
selection by running a \emph{Discovery Request procedure} with the
\gls{NRF}~\cite{3gpp.29.510}. We enhance client selection in \gls{NWDA}
by allowing the \gls{NWDAF} to decide the selection strategy and to
select \glspl{NF} by using metadata
(see~\Cref{registration-of-support-for-service}) or \glspl{KPI} via the
\gls{NWDA} \emph{Data Collection procedure}~\cite{3gpp.23.288}.

\hypertarget{local-model-update-retrieval}{%
\subsection{Local model update
retrieval}\label{local-model-update-retrieval}}

We integrate sending the local updates with the \gls{NWDA} analytics
subscribe pattern~\cite{3gpp.23.288}. In this way the \gls{NWDAF} can
trigger the \gls{NF} to start training of the \gls{ML} model on local
data. When the training is complete, the \gls{NF} sends the local model
update to the \gls{NWDAF}.

\hypertarget{global-model-update}{%
\subsection{Global model update}\label{global-model-update}}

The \gls{NWDAF} will aggregate all retrieved local model updates and
update the global model. The updated global model is sent to \glspl{NF}
that registered to receive global model updates.

Aggregation of the local model updates is done using a weighted average.
The weights depends on the number of local datapoints used in the
training round and therefore the local model update also need to contain
the number of datapoints.

\glspl{NF} can stop receiving global model updates from the \gls{NWDAF}
by using the analytics unsubscribe procedure\cite{3gpp.23.288}, for
example when the \gls{ML} performance of the global model is
sufficiently good.

\hypertarget{improving-privacy-further-using}{%
\section{\texorpdfstring{Improving Privacy further using
\gls{MPC}}{Improving Privacy further using }}\label{improving-privacy-further-using}}

In this section we present our second main contribution~--- a
privacy-enhancing scheme for \gls{FL} in the 5G~\gls{NWDA} framework inspired
by~\cite{Bonawitz}.

\subsection{Residual threats}

Even when \gls{FL} is used, observers may learn sensitive
information from the updates themselves.
That is, \glspl{NF} should not reveal their local updates
to the \gls{NWDAF}, nor to other \glspl{NF} or other parties.
Furthermore, \gls{FL} introduces a new type of sensitive
data into the system --- the number of datapoints used by each \gls{NF}. This
may leak information about the actual data. Therefore, our scheme also
protects the number of datapoints of each \gls{NF}.

Our scheme ensures that even if \glspl{NF} collude with the \gls{NWDAF} to
reveal the update of another \gls{NF}, they will fail.

It is still possible that some property of the inputs is deducible solely
from the output of the computation, i.e.,\ the global \gls{ML}
model~\cite{DBLP:conf/ccs/HitajAP17,FLleaks}.

\subsection{Privacy Through \acrlong{MPC}}

\begin{figure}
    \centering
    \begin{adjustbox}{width=.8\linewidth}%
        \begin{tikzpicture}[node distance=4cm,
            every node/.style={font=\large},
            object/.style={
                font=\Large
            },
            message/.style={
                font=\Large
            },
            phaselabel/.style={
                font=\bfseries\Large
            },
            descr/.style={
                    fill=white,
                    inner sep=2.5pt
                },
            comment/.style={
                    rounded corners=1mm, fill=EricssonYellow,
                    font=\sffamily\linespread{0.8}\selectfont,
                    inner sep=1mm,%
                    outer sep=1mm,%
                    minimum height=5mm,%
                    align=center,
                    text width=2.5cm},
            comment-right/.style={comment, right=0, signal, signal to=west },
            comment-left/.style={comment, left=.2, signal, signal to=east},
            connector/.style={
                font=\large
            },
            rectangle connector/.style={
                rounded corners=.25cm,
                connector,
                to path={(\tikztostart) -- ++(#1,0pt) \tikztonodes |- (\tikztotarget) },
                pos=0.5
            },
            rectangle connector/.default=-2cm,
            straight connector/.style={
                connector,
                to path=--(\tikztotarget) \tikztonodes
            }]

    \tikzstyle{object} = [rectangle, solid, minimum width=1.5cm, minimum height=.5cm,text centered, fill=EricssonGray5]
    \tikzstyle{lifeline} = [densely dotted, draw=black]
    \tikzstyle{phase} = [dashed, thick, draw=EricssonGreen, rounded corners=.5mm]
    \tikzstyle{arrow} = [thick,->,>=latex]
    \tikzstyle{message} = [outer sep=0pt, inner sep=0pt, font=\large]
    \tikzstyle{return} = [dotted]
    \tikzstyle{label} = [above, midway]
    \tikzstyle{phaselabel} = [below=.25cm, right, font=\large\bfseries]
    \tikzstyle{flowline} = [line width=4mm, EricssonRed, opacity=0.1, rounded corners=.25cm]
    \tikzstyle{mybrace} = [thick, decoration={calligraphic brace,raise=20pt, amplitude=3pt},decorate]

    \pgfmathparse{\colorname[0]};
    \edef\nfzerocolor{\pgfmathresult}

    \pgfmathparse{\colorname[1]};
    \edef\nfonecolor{\pgfmathresult}

    \draw node (NWDAF) at (0, 1) [object] {NWDAF} [lifeline] (NWDAF) -- ++(0,-14);
    \draw node (NF 0) [object, text=white, right of=NWDAF, fill=\nfzerocolor] {$\tNF{0}$} [lifeline] (NF 0) -- ++(0,-14);
    \draw node (NF 1) [object, text=white, right of=NF 0, fill=\nfonecolor] {$\tNF{1}$} [lifeline] (NF 1) -- ++(0,-14);
    \draw node (PKI) [object, right of=NF 1] {PKI} [lifeline] (PKI) -- ++(0,-14);


    \draw node [phaselabel] at  (-1, .5) {Session Initialization};
    \draw [phase] (-1, .5) rectangle (12.75, -9.5);

    \draw node [message] (in1) at (NWDAF |- 10,-0.5) {} [arrow] (in1) -- (in1 -|
    NF 1) node [label] {Key Setup Request $\left(1, \selList\right)$};


    \draw node [message] (in2) at (NF 1 |- 10,-1.25) {} [arrow, return] (in2) --
    (in2 -| NWDAF) node [label] {1.} node [midway, below
    right=0.05cm and -1.5cm] {$1, \left[\left\{ g^{x_{1,0}}, \pNF{0}, \pNF{1}\right\}\right]$};

    \draw node [message] (in3) at (NWDAF |- 10,-2.75) {} [arrow] (in3) -- (in3 -|
    NF 0) node [label] {2. $\left(2, \left\{ g^{x_{1,0}}, \pNF{0}, \pNF{1}\right\}\right)$};

    \draw [rectangle connector=-2.25cm, arrow] (in2 -| NWDAF) to node[descr] {Forward} (in3);

    \draw node [message] (in4) at (NF 0 |- 10,-3.75) {} [arrow, return] (in4) --
    (in4 -| NWDAF) node [label] {2.} node [return, midway, below
    right=0.05cm and -1.5cm] {$2, \left[\left\{g^{y_{1,0}}, \Sig_0\left(g^{x_{0,1}},g^{y_{1,0}}\right), \Mac_K\left(\pNF{0}\right)\right\}\right]$};

    \draw node [message] (in5) at (NWDAF |- 10,-5.25) {} [arrow] (in5) --
    (in5 -| NF 1) node [label] {3.} node [midway, below
    right=0.05cm and -4cm] {$3, \left\{g^{y_{1,0}}, \Sig_0\left(g^{x_{0,1}},g^{y_{1,0}}\right), \Mac_K\left(\pNF{0}\right)\right\}$};

    \draw [rectangle connector=-2.25cm, arrow] (in4 -| NWDAF) to node[descr] {Forward} (in5);

    \draw node [message] (pki3) at (NF 1 |- 10,-5.5) {} [arrow] (pki3) -- (pki3 -| PKI) node [label] {Key for $\tNF{0}$?};
    \draw node [message] (pki4) at (NF 1 |- 10,-6) {} [arrow, return] (pki4 -| PKI) -- (pki4) node [label] {Pub $\tNF{0}$};

    \draw node [message] (in6) at (NF 1 |- 10,-6.25) {} [arrow, return] (in6) --
    (in6 -| NWDAF) node [label] {3.} node [return, midway, below
    right=0.05cm and -1.5cm] {$3, \left[\left\{\Sig_1\left(g^{x_{0,1}}, g^{y_{1,0}}\right), \Mac_K\left(\pNF{1}\right)\right\}\right]$};

    \draw node [message] (in7) at (NWDAF |- 10,-7.75) {} [arrow] (in7) --
    (in7 -| NF 0) node [label] {4.} node [midway, below
    right=0.05cm and -3.5cm] {$4, \left\{\Sig_1\left(g^{x_{0,1}},g^{y_{1,0}}\right), \Mac_K\left(\pNF{1}\right)\right\}$};

    \draw [rectangle connector=-2.25cm, arrow] (in6 -| NWDAF) to node[descr] {Forward} (in7);

    \draw node [message] (pki3) at (NF 0 |- 10,-8.75) {} [arrow] (pki3) -- (pki3 -| PKI) node [label] {Key for $\tNF{1}$?};
    \draw node [message] (pki4) at (NF 0 |- 10,-9.25) {} [arrow, return] (pki4 -| PKI) -- (pki4) node [label] {Pub $\tNF{1}$};

    \draw node [phaselabel] at  (-1, -9.75) {Aggregation};
    \draw [phase] (-1, -9.75) rectangle (12.75, -13.25);


    \draw node [message] (m1) at (NWDAF |- 10,-10.75) {} [arrow] (m1) -- (m1 -| NF 1) node [label] {MPC Input Request ($t$)};
    \draw node [message] (in8) at (NF 1 |- 10,-11.5) {} [arrow, return] (in8) -- (in8 -| NWDAF) node [label] {MPC Input Response ( $-m_{(0,1)} + v_1$)};

    \draw node [message] (m2) at (NWDAF |- 10,-12.25) {} [arrow] (m2) -- (m2 -| NF 0) node [label] {MPC Input Request ($t$)};
    \draw node [message] (in9) at (NF 0 |- 10,-13) {} [arrow, return] (in9) -- (in9 -| NWDAF) node [label] {MPC Input Response ( $m_{(0,1)} + v_0$)};








\end{tikzpicture}%
    \end{adjustbox}%
    \caption[Overview of \acrfull{MPC} in \acrfull{FL} context]{An overview of
    \acrfull{MPC} in \acrfull{FL} context. In this
        example only two \glspl{NF} are selected. The \gls{NWDAF} instructs
        $\tNF{1}$ to initiate a key exchange with all \glspl{NF} with
        lower position, in this case only $\tNF{0}$. Following the session initialization we run an aggregation where
        the constructed mask $m_{(0,1)}$ is added by $\tNF{0}$ and
        subtracted by $\tNF{1}$.
        The key $K$ is derived from the Diffie-Hellman secret using a \gls{PRF}
        and is part of SIGMA.
        A list is denoted by brackets, and a container with curly brackets.}
    \label{fig:mpcsequencediagram}
    \squeezeup
\end{figure}
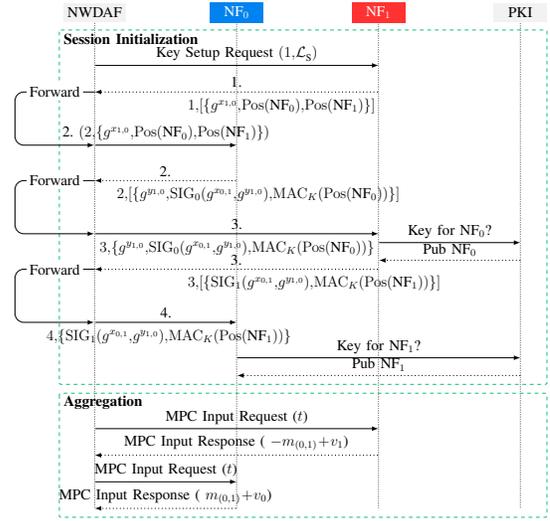

Our protocol consists of two parts, \emph{Session Initialization
} and \emph{Aggregation}.
The first part establishes a \emph{session} and pair-wise shared secrets
between the \glspl{NF}, and the second executes the
aggregation of local updates throughout a number of training rounds.
The local updates are protected by masks derived from the pair-wise shared
secrets.
An example of our scheme with two \glspl{NF} is depicted
in~\Cref{fig:mpcsequencediagram}.
%
%

\subsubsection{Preliminaries}
Session participants are an \gls{NWDAF} and a set of \glspl{NF} that we refer
to as the total population.
%
%
The total population is ordered according to some total order determined by
the \gls{NWDAF}, which remains fixed throughout the session.
When we refer to the position of an \gls{NF}, it is w.r.t. this order.
%

%
%
%

%
%

%
%
We assume that the \acrshort{PLMN} operator maintains a \gls{PKI}, in which all
\glspl{NF} are enrolled.
%
%
The \glspl{NF} are identified by their hostnames, which are unique within the
\gls{PKI}, and their private/public key pair represent their identity. All
participants have access to a fixed secure \gls{PRF} and a
fixed secure \gls{PRG} to compute the pair-wise shared secrets and masks,
see~\Cref{fig:mpcsequencediagram}.

%
%


\subsubsection{Session Initialization}
\label{sec:sessioninitialization}
The purpose of session initialization is to establish initialized
session states in the \gls{NWDAF} and all the \glspl{NF} in the total population.
%
%
%
%
%
The procedure is point-to-point between the \gls{NWDAF} and an \gls{NF} and
follows a request/response pattern according to~\cite[Section~4.6.1]{3gpp.29.501}.
\knote{Look for more references like this we can add to the paper to show the
tight connection to NWDA}
%
%
The procedure tunnels SIGMA key establishment~\cite{DBLP:conf/crypto/Krawczyk03}
messages between \glspl{NF} via the \gls{NWDAF}. 
SIGMA establishes pair-wise shared secrets between \glspl{NF}.
%
%
%
We allow caching of the pair-wise shared secrets between sessions.
A training round sequence number~$t$ ensures
fresh masks for each training round, even when
a pair-wise shared secret from a previous session is re-used.
The \gls{NWDAF} sets~$t$ to zero at the start of the session
and increases it by one for each round.
\glspl{NF} keep their own local replay counter $t_{\text{NF}}$, and abort
if the \gls{NWDAF} attempts to re-use a lower value for $t$.
The \gls{NWDAF} sends the first
message \emph{Key Setup Request} to all selected \glspl{NF}, which includes the
list $\selList$.
%

%
%

An $\tNF{k}$ acts as initiator of the SIGMA protocol execution if it has a
lower position than another \gls{NF}.
For each $\tNF{i}$, where ${i > k}$, $\tNF{k}$ creates a container that includes its
first SIGMA message.

The \gls{NWDAF} collects responses from all \glspl{NF}, before forwarding the
containers, in batch, to the correct \gls{NF}, based on the addresses on the
containers.
When an \gls{NF}, say $\tNF{k}$, receives a list of containers,
it creates the corresponding SIGMA response messages, packs them into containers
and returns them to the \gls{NWDAF}.
The remaining two SIGMA messages are exchanged similarly.
%
%
%

%
%
%
%
%
%
%

On completion of SIGMA, the session initialization is considered ready.
At this point, all pairs of \glspl{NF}, $\tNF{k}$
and $\tNF{k'}$, share a secret $\diffie_{k,k'}$ and associate the highest
seen training round sequence number with this Diffie-Hellman secret.
%
%

%
%

\subsubsection{Aggregation}

%
%
Local updates are hidden from the \gls{NWDAF} using masks,
which are shared using the secret sharing scheme
in~\Cref{sec:sessioninitialization}.
The masks cancel each other out during the execution of a secure sum
protocol, which is an optimization of~\cite[Protocol 1]{DBLP:conf/nordsec/KreitzDW10}.
%
%
Each \gls{NF} masks their local updates by independently and randomly
sample masks from
$\mathbb{Z}_R^{d}$ for some suitable $R$, where $d$ is the length of the
local update.
$R$ must be larger than any component of $\mLocUdtkt$, when the
component is interpreted as an unsigned integer.
The purpose of $R$ is to constrain the maximum size of components in the masked
local update.

The masks are combined with the local update $\mLocUdtkt$ by component-wise
integer addition modulo~$R$, where the components are considered as some fixed
integer encoding of their respective real-value.

The inverse is the component-wise additive inverse modulo~$R$. Finding the
inverse of a mask is trivial for someone who knows the mask and
infeasible for anyone else since masks are selected uniformly at random and are
independent.
%
%
Essentially, masking is encryption with a one-time pseudo-random pad.
%
%
%

%
Local updates can be large, therefore, so can masks.
To reduce communication overhead, we use an idea similar to~\cite{Bonawitz},
where masks are generated from the pair-wise secret shared between the
\gls{NF} adding the mask and the one canceling it.

\label{sec:aggregation}

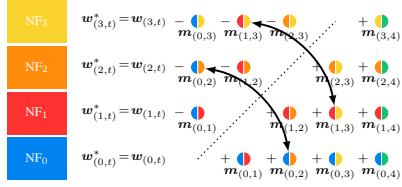
\begin{figure}
  \centering
  \setlength\abovecaptionskip{-.2\baselineskip}
  \begin{adjustbox}{width=.6\linewidth}%
    \begin{tikzpicture}

\def\ab{.65}

\tikzset{
  net node/.style = {rectangle, minimum width=2*\ab cm, inner sep=0.35 cm, fill=EricssonGray5},
}
\tikzset{font=\footnotesize}

\foreach \u in {0,...,3} {

    \pgfmathparse{\colorname[\u]};
    \edef\color{\pgfmathresult}

    \draw (-2.5,\u) node[net node, fill=\color, text=white] {$\tNF{\u}$};
    \draw (-.6,\u) node[black] {\(\mUdt{\u}{t}^* = \mUdt{\u}{t}\)};

    \pgfmathparse{max(0, \u - 1))};\edef\c{\pgfmathresult}

    \xifnum{\u > 0}{%
        \foreach \v in {0,...,\c} {
            \pgfmathparse{int(1 + mod(\v, 6))};
            \edef\xup{\pgfmathresult}

            \pgfmathparse{\colorname[\v]};
            \edef\color{\pgfmathresult}

            \pgfmathparse{\colorname[\u]};
            \edef\prevcolor{\pgfmathresult}

            \coordinate(down_\v_\u) at (\xup, \u);

            \draw[fill=\color, draw=white] ($ (down_\v_\u) + (0, 1ex) $) arc(90:270:1ex) -- cycle;
            \draw[fill=\prevcolor, draw=white] ($ (down_\v_\u) + (0, 1ex) $) arc(90:-90:1ex) -- cycle;

            \draw (down_\v_\u) node[below=0.5ex] {$\maskVec{\v}{\u}$} node[left=1ex] {$-$};
        }
    }

    \pgfmathparse{int(min(5, \u + 1)))};\edef\c{\pgfmathresult}

    \xifnum{\u < 4}{%
        \foreach \v in {\c,...,4} {
            \pgfmathparse{int(1 + mod(\v, 6))};
            \edef\xup{\pgfmathresult}

            \pgfmathparse{\colorname[\v]};
            \edef\color{\pgfmathresult}

            \pgfmathparse{\colorname[\u]};
            \edef\prevcolor{\pgfmathresult}

            \draw[fill=\prevcolor, draw=white] ($ (\xup, \u) + (0, 1ex) $) arc(90:270:1ex) -- cycle;
            \draw[fill=\color, draw=white] ($ (\xup, \u) + (0, 1ex) $) arc(90:-90:1ex) -- cycle;

            \coordinate(up_\u_\v) at (\xup, \u);
            \draw (\xup, \u) node[below=0.5ex] {$\maskVec{\u}{\v}$} node[left=1ex] {$+$};

        }
    }
}

\draw [semithick, shorten >=1ex,shorten <=1ex,<->] (up_0_2) to [bend right=35] (down_0_2);
\draw [semithick, shorten >=1ex,shorten <=1ex,<->] (up_1_3) to [bend right=35] (down_1_3);
\draw[thick, dotted] (1,0) -- (4,3);

\end{tikzpicture}%
  \end{adjustbox}%
  \caption[Cancellation of masks]{The 4 selected \glspl{NF} each add masks
  for every other \gls{NF}. The cancellation of
  masks \({\maskVec{0}{2} = \protect\sharedkey{0}{2}}\) and
  \({\maskVec{1}{3} = \protect\sharedkey{1}{3}}\)  are indicated by the arrows.
  Each mask above the diagonal is canceled out by a mask
  below the diagonal. The cancellations of masks
  do not affect the sum, so
  \(\sum_{i=0}^{3} \mUdt{i}{t}^* = \sum_{i=0}^{3} \mUdt{i}{t}\).}
  \label{fig:masks}
  \squeezeup
\end{figure}

\label{sect:locUpdate}
%
%
%
%
%
%
%
Consider masks as vectors of unsigned integers and let the position of an \gls{NF}
be $k$.
$\tNF{k}$ hides its local update~$\mLocUdtkt$ by adding all masks~$\mKI$ to it where
$k < i$, and subtracting all masks~$\mKI$ from it where $k > i$.
The masked version~$\mLocUdtkt^*$ of the local update is
\begin{displaymath}
    \mLocUdtkt^* = n_k \cdot \mLocUdtkt +
    \sum_{\substack{i \in \selList,\\ k < i}} \mKI -
    \sum_{\substack{i \in \selList,\\ k > i}} \mIK,
\end{displaymath}
where \(n_k\) is the number of datapoints for $\tNF{k}$ used in training round~$t$,
see~\Cref{fig:masks} for a visual representation of this.

As a last step, $\tNF{k}$ generates a separate set of masks to
hide the number of datapoints used for training.
These masks are added and subtracted in the same way as the masks for the
local updates, and the \gls{NWDAF}
adds them together and uses the result to scale the sum by the total number of
points~$n$ used during this
training round.
This protects the number of datapoints, which would otherwise potentially reveal
sensitive information about the local data.

\hypertarget{evaluation}{%
\section{Evaluation}\label{evaluation}}

\label{sec:evaluation} In this section we present our third main
contribution~--- an evaluation of the proposed privacy-preserving scheme
for \gls{FL} in 5G~\gls{NWDA}. An argument for the security of the
scheme can be found~in \Cref{app:security}.

\hypertarget{setting}{%
\subsection{Setting}\label{setting}}

Security protocols add overhead, and it is important to keep the ratio
between security overhead and protected data low.

5G datasets can be sensitive in a business sense so real associated
models are therefore also unpublished which makes it difficult to obtain
and use such a model for our analysis. Therefore,
in~\Cref{fig:usecases}, as a reading guide for this section we provide a
visual representation of the operating point (in terms of number of
\glspl{NF} and model size) for some use-cases. We consider the case
where \glspl{NF} are base stations deployed for one operator and provide
the estimated number of \glspl{NF} based on the number of deployed 4G
base stations for a small and a large operator~\cite{cellmapper}.

The size of the \gls{ML} model depends on the use-case, from very small
models in~\cite{DiazGonzalez1334598} to larger models. As use-cases
become more complex, and as the volume of data generated increases, the
needed size of models will also increase. We expect the operating point
for future 5G use-cases to end up in the green area seen
in~\Cref{fig:usecases}.

\hypertarget{communication-cost}{%
\subsection{Communication cost}\label{communication-cost}}

We use a communication cost metric where we include the total number of
bytes transmitted both in uplink and in downlink. Note that the lower
layer protocol overhead, from for example HTTP/2 and \acrshort{TLS}, is
not included in our analysis.

\begin{figure*}[!tb]
    \centering
    \setlength\abovecaptionskip{-.5\baselineskip}
    \begin{minipage}[t]{.232\linewidth}
        \centering
        \begin{adjustbox}{width=1\linewidth}\input{cost_per_client.tex}\end{adjustbox}%
        \caption{Comparison of communication costs after one~training round between session initialization messages and aggregation messages. We vary the fraction of \glspl{NF} selected $C$ out of \num{{}e5} \glspl{NF}. At $C \approx 0.4$ the cost of the session initialization
        messages exceed the aggregation cost.}
        \label{fig:cost_of_messages}
    \end{minipage}%
    \hfill%
    \begin{minipage}[t]{.232\linewidth}
        \centering
        \begin{adjustbox}{width=1\linewidth}
\begin{tikzpicture}
\path (-1.45cm,-2.69cm) rectangle (6.6cm, 3.55cm);

\definecolor{color0}{rgb}{0,0.407843137254902,0.749019607843137}
\definecolor{color1}{rgb}{0.8,0.156862745098039,0.156862745098039}

\begin{axis}[
legend cell align={left},
legend columns=2,
legend style={at={(0.5,-0.4)}, anchor=north, draw=none, fill=none},
tick align=outside,
tick pos=left,
x grid style={white!69.01960784313725!black},
xlabel={Training round},
xmin=1, xmax=25,
xtick style={color=black},
y grid style={white!69.01960784313725!black},
ylabel={Communication cost [\si{\mebi\byte}]},
ymin=0, ymax=50000000,
ytick style={color=black}
]
\addplot [semithick, color0]
table {%
0 0
1 3715307.81930542
2 7430615.63861084
3 11145923.4579163
4 14861231.2772217
5 18576539.0965271
6 22291846.9158325
7 26007154.7351379
8 29722462.5544434
9 33437770.3737488
10 37153078.1930542
11 40868386.0123596
12 44583693.831665
13 48299001.6509705
14 52014309.4702759
15 55729617.2895813
16 59444925.1088867
17 63160232.9281921
18 66875540.7474976
19 70590848.566803
20 74306156.3861084
21 78021464.2054138
22 81736772.0247192
23 85452079.8440247
24 89167387.6633301
};
\addlegendentry{Aggregation phase}
\addplot [semithick, color1]
table {%
0 3705615.32790756
1 6601643.5458823
2 8868048.50001247
3 10644789.7113402
4 12040714.9732204
5 13140476.2052223
6 14009907.9756536
7 14700210.4274798
8 15251202.377883
9 15693851.2833963
10 16052240.8168155
11 16345101.0696711
12 16586998.6046183
13 16789261.9700113
14 16960701.4810121
15 17108168.999909
16 17236993.2823659
17 17351318.550162
18 17454367.8022939
19 17548647.5944151
20 17636107.2976709
21 17718262.955743
22 17796293.6096011
23 17871116.2101383
24 17943443.8784096
};
\addlegendentry{Initialization phase}
\addplot [semithick, color1, dashed]
table {%
0 3705615.32790756
1 7411230.65581512
2 11116845.9837227
3 14822461.3116302
4 18528076.6395378
5 22233691.9674454
6 25939307.2953529
7 29644922.6232605
8 33350537.9511681
9 37056153.2790756
10 40761768.6069832
11 44467383.9348907
12 48172999.2627983
13 51878614.5907059
14 55584229.9186134
15 59289845.246521
16 62995460.5744286
17 66701075.9023361
18 70406691.2302437
19 74112306.5581512
20 77817921.8860588
21 81523537.2139664
22 85229152.5418739
23 88934767.8697815
24 92640383.1976891
};
\addlegendentry{Initialization phase\\ w/o memory}
\end{axis}

\end{tikzpicture}\end{adjustbox}%
        \caption{Comparison of communication costs per training round between
        session initialization messages and aggregation messages. The number of key establishments are reduced as the number of training rounds
        increases. The fraction
        of \glspl{NF}
        selected $C$ is from the intersection point in~\Cref{fig:cost_of_messages}.}
        \label{fig:cost_of_messages_epochs}
    \end{minipage}%
\hspace{0.2in}%
\begin{minipage}[t]{.232\linewidth}
    \centering
    \begin{adjustbox}{width=1\linewidth}\begin{tikzpicture}
\path (-1.45cm,-2.69cm) rectangle (6.6cm, 3.55cm);
\tikzset{font=\footnotesize}
\definecolor{color0}{rgb}{0,0.407843137254902,0.749019607843137}
\definecolor{color1}{rgb}{0.8,0.156862745098039,0.156862745098039}
\definecolor{color2}{rgb}{0.8,0.43921568627451,0.0274509803921569}
\definecolor{color3}{rgb}{0.776470588235294,0.658823529411765,0.137254901960784}
\definecolor{color4}{rgb}{0.0470588235294118,0.607843137254902,0.356862745098039}

\begin{axis}[
legend cell align={left},
legend columns=3,
legend style={at={(0.5,-0.35)}, anchor=north, draw=none},
tick align=outside,
tick pos=left,
x grid style={white!69.01960784313725!black},
xlabel={\gls{ML} model size [\si{\byte}]},
xmin=0, xmax=1e9,
xtick style={color=black},
y grid style={white!69.01960784313725!black},
ylabel={Number of \gls{NF}},
ymin=0, ymax=1e6,
ytick style={color=black},
xmode=log,
ymode=log
]

\addplot+[EricssonBlue, mark=*,mark options={scale=1, fill=EricssonBlue}, font=\small, nodes near coords,only marks,
point meta=explicit symbolic, every node near coord/.style={anchor=south west}]
table[meta=label] {
Depth Breadth label
1568 120 {LSTM \cite{DiazGonzalez1334598}}
};

\filldraw[white, even odd rule,inner color=EricssonGreen!80,outer color=white] (axis cs:1e4,1000) rectangle (axis cs:1e7,1e5);

\path [draw=black, semithick, dash pattern=on 1.5pt off 2.4749999999999996pt]
(axis cs:0,4237)
--(axis cs:1e9,4237);

\path [draw=black, semithick, dash pattern=on 1.5pt off 2.4749999999999996pt]
(axis cs:0,67067)
--(axis cs:1e9,67067);

\path [draw=EricssonBlue, semithick]
(axis cs:0,1e5)
--(axis cs:1e9,1e5);

\node at (axis cs:2e3,1e5)[
  scale=0.6,
  anchor=south west,
  text=EricssonBlue,
  rotate=0.0,
  font=\large
]{Bonawitz \emph{et al.}~\cite{Bonawitz}};

\node at (axis cs:2e3,4237)[
  scale=0.6,
  anchor=south west,
  text=black,
  rotate=0.0,
  font=\large
]{Small Operator};

\node at (axis cs:2e3,67067)[
  scale=0.6,
  anchor=north west,
  text=black,
  rotate=0.0,
  font=\large
]{Large Operator};



\path [draw=black, semithick, dash pattern=on 1.5pt off 2.4749999999999996pt]
(axis cs:241376928,0)
--(axis cs:241376928,1e66);



\node at (axis cs:241376928,4e5)[
  scale=0.6,
  anchor=south east,
  text=black,
  rotate=0.0,
  font=\large
]{ResNet-152 \cite{DBLP:conf/cvpr/HeZRS16}};

\end{axis}

\end{tikzpicture}\end{adjustbox}%
    \caption{A guide to understanding the costs and trade-offs we provide ball-park estimates of number of \glspl{NF} and model sizes for a few use-cases. We expect the operating point for future 5G use-cases to end up in the green area.}
    \label{fig:usecases}
\end{minipage}%
\hfill%
\begin{minipage}[t]{.232\linewidth}
    \centering
    \begin{adjustbox}{width=1\linewidth}\input{cost_bonawitz.tex}\end{adjustbox}%
    \caption{\gls{DEF} compared to~\cite{Bonawitz} in the fully malicious case. Line type represents the solution, color represents the number of \glspl{NF}. The decreasing \gls{DEF} with increasing training rounds of our scheme is shown for the first training round and for the case when all \glspl{NF} have been selected at least once.}
    \label{fig:cost_comparison}
\end{minipage}
\squeezeup
\end{figure*}

\hypertarget{session-initialization-security-related-communication}{%
\subsubsection{Session initialization: security related
communication}\label{session-initialization-security-related-communication}}

\hypertarget{first-round}{%
\paragraph{First round}\label{first-round}}

In~\Cref{fig:mpcsequencediagram} we see an overview of the messages in
the \emph{Session Initialization}. The \gls{NWDAF} sends a list of
hostnames \(\selList\) to each of the selected \(K_s\)~\glspl{NF}. Each
\gls{NF} will respond to this message with a list of transparent
containers, one for each \gls{NF} that has a lower index than the
initiating \gls{NF}, in total \(\binom{K_s}{2}\) such containers. In
this way, we calculate the total communication cost of messages in
session initialization and compare between the session initialization
and aggregation communication cost in~\Cref{fig:cost_of_messages}.

\hypertarget{subsequent-training-rounds}{%
\paragraph{Subsequent training
rounds}\label{subsequent-training-rounds}}

The probability that selected \glspl{NF}, \(\tNF{i}\) and \(\tNF{j}\),
need to exchange secrets in round~\(t\) is \begin{displaymath}
P_\text{$\tNF{i}$ key exchange with $\tNF{j}$} = \left(1 - \frac{K_s}{K}\left[\frac{K_s-1}
{K-1}\right]\right)^t.
\end{displaymath} As \(t\) goes to infinity, this probability goes to
\num{0}, as seen in~\Cref{fig:cost_of_messages_epochs} where the
communication cost growth for the session initialization drops off. Note
that an \gls{NF} may already have a certificate of another \gls{NF} that
it obtained for some other reason.

Communication cost in the session initialization is \BigO{K_s^2}. Note
that this cost is heavily influenced by the size of the hostname list.
Each \gls{NF} is assigned a \SI{12}{\byte} globally unique
identifier~\cite{3gpp.29.571}. Adding a domain name, the hostnames we
use are \SI{30}{\byte} of the form
\url{ gNB-382A3F47.myran.example.com}. Other security parameters are
chosen to match \SI{128}{\bit} security.

\hypertarget{aggregation-model-related-communication}{%
\subsubsection{\texorpdfstring{Aggregation: \gls{ML} model related
communication}{Aggregation:  model related communication}}\label{aggregation-model-related-communication}}

In each training round each selected \gls{NF} sends a local model update
to the \gls{NWDAF}, see~\Cref {fig:flsequencediagram}. The \gls{NWDAF}
performs aggregation and sends the updated global model to all
registered \glspl{NF}. The total cost of sending the model of size \(d\)
in this step depends on \(K\) and on \(K_s\). Each \gls{NF} has a
communication cost of \BigO{d}, so the total aggregation communication
cost in each round is~\BigO{dK}.

\hypertarget{protocol-overhead}{%
\subsubsection{Protocol overhead}\label{protocol-overhead}}

We compare the total size of all security related messages and model
related messages to the protocol without security, i.e.~standard
\gls{FL}. We call this overhead \gls{DEF}. The \gls{DEF} from security
related messages increases with increasing number of \glspl{NF}. We
expect future \gls{FL} use-cases in 5G to require large models and
relatively small number of \glspl{NF}.
\knote{These use-cases in 5G (presumably the same thing as "the 5G use-cases" seems to be very broad: any use-case for FL in 5G. Is that the intention?}

In~\Cref{fig:cost_comparison} we compare the \gls{DEF} of our scheme to
that of~\cite{Bonawitz} in their fully malicious case. This comparison
is done for the cases where the number of \glspl{NF} match the number of
base stations from a small and large operators. We vary the model size.
As seen in~\Cref{fig:cost_of_messages_epochs} the \gls{DEF} for our
scheme is reduced with increasing~\(t\). In \Cref{fig:cost_comparison}
we plot the \gls{DEF} for our scheme when all \glspl{NF} have been
selected at least once~--- there is a similar but much small effect
for~\cite{Bonawitz} but this is omitted for clarity.

\hypertarget{computation-cost}{%
\subsection{Computation cost}\label{computation-cost}}

\hypertarget{session-initialization-computation}{%
\subsubsection{Session initialization
computation}\label{session-initialization-computation}}

Each \gls{NF} will at most need to do \(K_s-1\) key establishments in
one training round and need to expand the seed to a full mask for every
other \gls{NF}. The mask length depends on the model size. The resulting
session initialization computation cost per \gls{NF} is~\BigO{K_sd}.

The session initialization computation cost for the \gls{NWDAF} depends
on the number of selected \glspl{NF} \(K_s\), and is \BigO{K_s}.

\hypertarget{aggregation-computation}{%
\subsubsection{Aggregation computation}\label{aggregation-computation}}

The \gls{NWDAF} is unaware of any masks, and simply performs aggregation
of \(K_s\) local model updates. The resulting session initialization
computation cost for the \gls{NWDAF} is~\BigO{K_sd}. The \gls{ML}
computation costs for an \gls{NF} (training and inference) are out of
scope of this \manuscript{}.

\hypertarget{storage-costs}{%
\subsection{Storage costs}\label{storage-costs}}

\hypertarget{session-initialization-storage}{%
\subsubsection{Session initialization
storage}\label{session-initialization-storage}}

The largest storage needed is for \(K-1\) session keys, \(K\)
certificates, \(K\) hostnames, \num{1}~private key and \num{1}~training
round sequence number. We can trade storage for communication by only
storing the certificates and session keys for the \glspl{NF} that are
selected in the current training round. The storage needed in this case
then depends on \(K_s\).\mnote{Trade-off paragraph needed?} The storage
needed for the \gls{NWDAF} is for \(K\) hostnames and \num{1}~training
round sequence number.

\hypertarget{aggregation-storage}{%
\subsubsection{Aggregation storage}\label{aggregation-storage}}

Each participating \gls{NF} need to store the global model, not counting
temporary storage needed during training. The model related storage for
each \gls{NF} is \BigO{d} and the state related storage is \BigO{K_s}.
The model related storage for the \gls{NWDAF} is \BigO{d}.

\hypertarget{discussion-and-related-work}{%
\section{Discussion and related
work}\label{discussion-and-related-work}}

\hypertarget{related-work}{%
\subsection{Related Work}\label{related-work}}

Our scheme is inspired by~Bonawitz
\emph{et al.}~\cite{DBLP:journals/corr/abs-1902-01046} in which the
authors discuss a practical implementation of \gls{FL} including
security. Their security aspects are further developed
in~\cite{Bonawitz}. They target mobile devices with no pre-established
security relations and where group membership is volatile. They overcome
this volatility by additional functionality in their scheme. However, as
we assume that \glspl{NF} will have a much lower drop rate than mobile
phones, we avoid their robustness-improving additions. We also make use
of the fact that \glspl{NF} already are part of a common PKI to reduce
complexity. This excludes use-cases with more than one mobile operator,
such as~\cite{DBLP:journals/corr/abs-1906-09248,EricssonReviewFL}, and
we leave this as future work.

A parameter server may detect malicious or malfunctioning clients based
on the information in the local updates.
\cite{Damaskinos2019}~implements a robust Byzantine-resilient
aggregation method. Unfortunately, such methods fail when \gls{MPC} is
used, because they need access to the local updates of the \gls{FL}
clients.

\cite{DBLP:conf/globecom/ZhangCYD19}~proposed to encrypt local updates
using Paillier homomorphic cryptosystem which can be more efficient in
the initialization phase, but don't evaluate this. They show that
parameter server complexity increases, and that \gls{ML} performance is
lower. They do not embed their protocol in any particular system.

Although \gls{MPC} is applied, the global model may leak information.
Differential privacy could be used, but those schemes need further work
before they can be practically
applied~\cite{DBLP:journals/corr/abs-1912-04977}.

\hypertarget{conclusion}{%
\section{Conclusion}\label{conclusion}}

\gls{ML} is becoming an essential technology for optimizing mobile
networks. This has lead to an increased collection and processing of
data that may leak sensitive information. Consequently, mechanisms to
protect the business sensitive information and end-users' privacy are
needed.

We devised a scheme for end-user privacy protection and demonstrated how
to integrate it in the 5G \gls{SBA} and \gls{NWDA} architecture. The
scheme was evaluated in terms of computational and communication cost.
We explore the security of our scheme
in~\cite[Appendix A]{ourlongerpreprint}.

We found that the communication overhead, \gls{DEF}, depends on the
client fraction \(C\), the size of the \gls{ML} model, the number of
\glspl{NF} and the training round~\(t\). For the use-cases we envision,
as well as for potential future use-cases, we showed that the overhead
of our scheme is smaller than that of~\cite{Bonawitz}. Our gain stemmed
from relaxed reliability constraints and re-use of existing telecom
infrastructure, such as \gls{PKI}. However, we see an opportunity to
further improve our scheme in terms of communication overhead and to use
our \gls{NWDA} integration to improve bias and fairness.

Even though it is known that sensitive information may still leak even
when \gls{FL} and \gls{MPC} are properly applied, our scheme
significantly improves privacy. Because it is available and much simpler
to apply in practice, in comparison to differential privacy, we believe
it would be beneficial to deploy a scheme such as ours.

\section*{Acknowledgment}\label{acknowledgments}

This work was partially supported by the Wallenberg AI, Autonomous
Systems and Software Program (WASP) funded by the Knut and Alice
Wallenberg Foundation.

We would especially like to thank Prof.~Patric Jensfelt, Prof.~Mads Dam
and Dr.~Rickard Cöster and all anonymous reviewers for their invaluable
input. \footnotesize

\bibliographystyle{IEEEtran}
\balance
\bibliography{references/refs}

\balance
\normalsize
\clearpage
\appendix
\section{Appendix}
\subsection{Security justification}\label{app:security}

Although we have explained the security purpose for introducing functionality
throughout the paper, we now give a brief argument for the security of the
compound scheme.
The correctness of the protocol can be seen from the description of the protocol
itself above, and we will not consider it further.

The security goal of the scheme is to ensure that the \gls{NWDAF}
only knows its own input to the computation and the final result, i.e., the
updated model.
The initialization phase is run once and the masked-based secure sum protocol
is run each round to compute a new updated model.
Our argument that this is secure can be divided into the following two claims.

\emph{Claim 1:} The mask-based secure sum protocol fulfills the security goal
assuming the masks are uniformly and randomly selected and assuming each pair
of \glspl{NF} share a mask and its inverse (we call these two pair-wise masks
below), known only to that \gls{NF}-pair.

\emph{Claim 2:} The masks are uniformly and randomly selected in each round
assuming SIGMA is a secure key establishment protocol, that the \gls{PRF},
\gls{PRG}, signature scheme and MAC are secure according to standard definitions.

To justify \emph{Claim 1} we argue as follows.
For each other \gls{NF}, an \gls{NF} adds the mask it shares with that \gls{NF}
to the local update (or the inverse of the mask depending on their relative
positions in the \gls{NF} order).
Adding a uniformly random mask to a local update using modular addition results
in a uniform distribution.
No-one except that pair of \glspl{NF} can hence distinguish the masked local
update from a random value.
For each pair of \glspl{NF}, the \gls{NF} adds a mask known only to that pair.
This means that given the total sum of all the masks and the local update contains
at least $|\selList| - 2$ masks not known to any given \gls{NF}.
Consequently, at least $|\selList| - 2$ need to collude to unmask a masked local
update.
Adding two masked local updates together will provide the sum of those local
updates still masked by the remaining masks.
As long as at least one mask remains in the sum, the sum cannot be unconcealed.
Once all local updates are added together, the result is their sum and all masks
are canceled.
At this point the \gls{NWDAF} knows the output of the computation, but has not
been able to unconceal any of the inputs, which is what we claimed.

To justify \emph{Claim 2} we argue as follows.
In the initialization phase, each \gls{NF} establishes a pair-wise secret with
each other \gls{NF} using the SIGMA key establishment protocol.
Further, the signature scheme, MAC and MAC-key generation via the \gls{PRF},
on which SIGMA relies, are secure by assumption.
SIGMA is secure in the CK-model~\cite{DBLP:conf/crypto/Krawczyk03},
meaning that the established shared secret is indistinguishable from a randomly
selected element from the underlying Diffie-Hellman group.
We can therefore assume that the pair-wise shared secret is indistinguishable
from random to anyone else than the pair of \glspl{NF} and that it is mutually
authenticated.

In each round, each \gls{NF} verifies that $t$ has not been used earlier, and we
therefore can assume it is fresh for all \glspl{NF} in all runs of the protocol.
For simplicity, we assume that an \gls{NF} that detects a re-used $t$ value
stops execution, at which point the entire round of the protocol fails to
execute.
Note that even in that case, only \glspl{NF} which obtain a fresh $t$ value
would continue execution, so all \glspl{NF} can be assumed to use a fresh $t$
value to generate output in the protocol.

The pair-wise masks are generated from the pair-wise shared secret, which may
be the same for more than one training round.
However, the value $t$, which is guaranteed to be fresh for each training round,
is also used as input to the mask-generating \gls{PRG}.
Because the \gls{PRG} is secure, its output is indistinguishable from a
uniformly randomly selected string given that the input $g^{xy}$ obtained
from SIGMA is secret.
To conclude, because $t$ is fresh, the pair-wise masks are uniformly random,
known only to the \gls{NF}-pair, and they are secret and fresh for each round.

We note that the order of the \glspl{NF} affects two aspects.
First, the order determines which \gls{NF} acts as initiator and which one acts as
responder for the SIGMA exchange between each pair.
SIGMA is secure regardless which part takes which role, and no \gls{NF} will
continue execution of the scheme unless it has run SIGMA with each \gls{NF} in
the order.
So, the \gls{NWDAF} does not gain anything by selecting a certain order in this
respect.

Second, the order determines which \gls{NF} computes a mask and which \gls{NF}
computes the inverse of a mask.
By the symmetry of the masks, it is irrelevant which \gls{NF} generates the
mask.
Since no \gls{NF} will continue execution unless it has a pair-wise mask with
all other \glspl{NF} in the pair, the \gls{NWDAF} gain nothing by selecting
a certain order.

The secure sum protocol is considered meeting the security goal even
in the degenerate case where the $|\selList| = 1$.
In that case, the \gls{NWDAF} would in fact learn the local update of the
single participating \gls{NF}.
This can be prevented by adding a rule in the scheme that \glspl{NF} shall
terminate the execution if the size $\selList$ is less than some threshold value.
It may be useful to set this threshold to a larger value than one to reduce
the effects of outliers.
%


\end{document}